\documentclass[fleqn,12pt,twoside]{article}
\usepackage[headings]{espcrc1}

\readRCS
$Id: espcrc1.tex,v 1.2 2004/02/24 11:22:11 spepping Exp $
\usepackage{graphicx}
\usepackage[figuresright]{rotating}

\newcommand{\AmS}{{\protect\the\textfont2
  A\kern-.1667em\lower.5ex\hbox{M}\kern-.125emS}}

\title{Multicritical point of spin glasses
\footnote{Dedicated to Prof. A. Nihat Berker on the occasion
of his sixtieth birthday.}
}

\author{Hidetoshi Nishimori \address[HN]{Department of Physics, Tokyo Institute of Technology, \\
Oh-okayama, Meguro-ku, Tokyo 152-8551, Japan}
, and Masayuki Ohzeki\addressmark}
       
\runtitle{Multicritical point of spin glasses}
\runauthor{H. Nishimori and M. Ohzeki}

\begin{document}

\maketitle

\begin{abstract}
We present a theoretical framework to accurately calculate the location
of the multicritical point in the phase diagram of spin glasses.
The result shows excellent agreement with numerical estimates.
The basic idea is a combination of the duality relation,
the replica method, and the gauge symmetry.
An additional element of the renormalization group, in particular in the context
of hierarchical lattices, leads to impressive improvements of the predictions.
\end{abstract}

\section{INTRODUCTION}

Identification of the precise location of the multicritical point is
an important theoretical challenge in the physics of spin glasses
not only because of its mathematical interest but also for the practical
purpose of reliable analyses of numerical data.
The method of duality is a standard tool to derive th exact location
of a critical point in pure ferromagnetic systems in two dimensions.
However, the existence of randomness in spin glasses hampers
a direct application of the duality.

We have nevertheless developed a theory to achieve the goal by
using the combination of the replica method, the duality applied to
the replicated  system, the gauge symmetry,
and the renormalization group \cite{NN,MNN,TSN,Nstat,ONB,Ohzeki}.
The result shows excellent agreement with numerical estimates.
The analysis on hierarchical lattices plays a crucial role
in the development of the theory, in particular in the introduction
of the renormalization group, by which systematic improvements
can be achieved.

\section{MULTICRITICAL POINT}
\begin{figure}[tbp]
\begin{center}
\includegraphics[width=50mm]{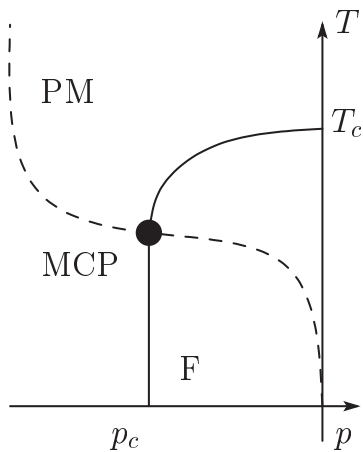}
\end{center}
\caption{{\protect\small A typical phase diagram of the $\pm J$ Ising model in two dimensions.
The multicritical point (MCP) is described by a black dot and
the Nishimori line is drawn  dashed.}}
\label{fig1}
\end{figure}

Let us consider the $\pm J$ Ising model defined by the Hamiltonian,
\begin{equation}
H = - \sum_{\langle ij \rangle} J_{ij} \sigma_{i}\sigma_{j},
\end{equation}
where $\sigma_i$ is the Ising spin and $J_{ij}$ denotes the quenched random coupling.
The sign of $J_{ij}$, i.e. $J_{ij}/J=\tau_{ij}$, follows the distribution
\begin{eqnarray}
P(\tau_{ij}) &=& p \delta(1-\tau_{ij}) + (1-p) \delta(1+\tau_{ij})
\nonumber\\
&=&
\frac{\exp(K_p \tau_{ij})}{2\cosh K_p} \left\{\delta(1-\tau_{ij}) + \delta(1+\tau_{ij})\right\},
\end{eqnarray}
where $\exp(-2K_p) = (1-p)/p$.
The multicritical point is believed to lie on the Nishimori line (NL)
defined by $K_p=\beta J$, where $\beta$ is the inverse temperature.
See Fig. \ref{fig1}.
The restriction to the NL simplifies the problem due to the gauge symmetry \cite{HN81,HNbook}.

According to the initial theory that uses the replica method, duality and gauge symmetry
\cite{NN,MNN,TSN,Nstat}, the value of $p_c$ for the multicritical
point satisfies
\begin{equation}
 H(p_c) = \frac{1}{2},
 \label{conjecture}
\end{equation}
where $H(p)$ is the binary entropy, $-p\log_2 p-(1-p)\log_2(1-p)$,
for self-dual lattices.
Equation (\ref{conjecture}) is solved to give $p_c=0.8900$,
which is in reasonable agreement with numerical estimates.
The theory has also been extended to a pair of mutually dual lattices with $p_{c1}$
and $p_{c2}$ for respective multicritical points.  The result is
\begin{equation}
H(p_{c1})+H(p_{c2})=1.
\label{Hpc}
\end{equation}
Hinczewski and Berker, however, found $H(p_1)+H(p_2)=1.0172,0.9829,0.9911$
for three pairs of mutually dual hierarchical lattices \cite{HB}.
Their values are correct to the decimal points shown above as one can carry out
numerically exact renormalization group calculations on hierarchical lattices.
Thus Eq. (\ref{Hpc}) is a good approximation but not quite exact,
at least for hierarchical lattices.

\section{REPLICA AND DUALITY}
Let us give a very brief summary of the theory that leads to
Eqs. (\ref{conjecture}) and (\ref{Hpc}).
We generalize the usual duality argument to the $n$-replicated $\pm J$ Ising model.

We define the edge Boltzmann factor $x_k ~(k = 0, 1, \cdots, n)$, which represents
the configuration-averaged Boltzmann factor for interacting spins with $k$ antiparallel
spin pairs among $n$ nearest-neighbour pairs for a bond (edge).
The duality gives the following relationship between the partition functions
on the original and dual lattices with different values of the edge Boltzmann factors
\begin{equation}
Z_n(x_0,x_1,\cdots,x_n) = Z_n(x^*_0,x^*_1,\cdots,x^*_n),  \label{PF2}
\end{equation}
where we have assumed self duality of the lattice in that
both sides share the same function $Z_n$.
The dual edge Boltzmann factors $x_k^*$ are defined by the discrete multiple Fourier
transforms of the original edge Boltzmann factors, which are simple combinations of
plus and minus of the original Boltzmann factors in the case of Ising spins.

It turns out useful to focus our attention to the  principal Boltzmann factors $x_0$ and $x^*_0$,
which are the most important elements of the theory.
Their explicit forms are
\begin{equation}
x_0(K,K_p) = \frac{\cosh \left(nK + K_p \right)}{\cosh K_p},\quad
x^*_0(K,K_p) = \left( \sqrt{2} \cosh K \right)^n,
\end{equation}
where $K=\beta J$.
We extract these principal Boltzmann factors from the partition functions in Eq. (\ref{PF2}),
which amounts to measuring the energy from the all-parallel spin configuration.
Then, using the normalized edge Boltzmann factors $u_j = x_j/x_0$ and $u^*_j = x^*_j/x^*_0$,
we have
\begin{equation}
{x_0(K,K_p)}^{N_B}z_n(u_1,u_2,\cdots,u_n) = {x^*_0(K,K_p)}^{N_B}z_n(u^*_1,u^*_2,\cdots,u^*_n),
 \label{PF1}
\end{equation}
where $z_n(u_1,\cdots)$ and $z_n(u^*_1,\cdots)$ are defined as $Z_n/x^{N_B}_0$
and $Z_n/(x^{*}_0)^{N_B}$ and $N_B$ is the number of bonds.

\begin{figure}
\begin{center}
\includegraphics[width=60mm]{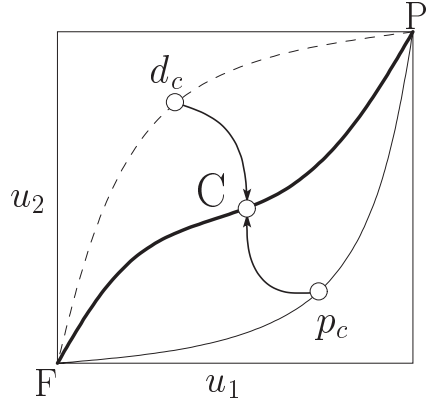}
\end{center}  
\caption{
A schematic picture of the renormalization flow and the duality
for the replicated $\pm J$ Ising model.}
\label{Trajectory}
\end{figure}

We now restrict ourselves to the NL, $K=K_p$.
Figure \ref{Trajectory} shows the relationship between the curves
$(u_1(K),u_2(K),\cdots,u_n(K))$ (the thin curve) and $(u^*_1(K),u^*_2(K),\cdots,u^*_n(K))$
(the dashed curve).
The arrows emanating from both curves represent the renormalization flows
toward the fixed point C.

The ordinary duality argument identifies the critical point under
the assumption of a unique phase transition.
We can obtain the critical point as the fixed point of the duality
transformation using the fact that the partition function is a single-variable function.
In other words, the thin curve would overlap with the dashed line for such a case.

In the present random case, on the other hand, since $z_n$ is a multivariable
function, there is no fixed point of the duality in the strict sense
which satisfies $n$ conditions simultaneously,
$u_1(K)=u^*_1(K),u_2(K)=u^*_2(K),\cdots,u_n(K)=u^*_n(K)$.
This is in sharp contrast to the non-random Ising model.
We nevertheless assume that $x_0(K,K)=x_0^*(K,K)$ may give the precise location
of the multicritical point because, when the number of variables of $z_n$ in
Eq. (\ref{PF1}) is unity ($n=1$), the fixed point condition $u_1=u^*_1$
implies $x_0=x^{*}_0$.
This relation, in the limit of $n\to 0$ in the spirit of the replica method,
leads to Eq. (\ref{conjecture}).
A straightforward generalization to mutual dual cases gives Eq. (\ref{Hpc}).

\section{RENORMALIZATION GROUP ON HIERARCHICAL LATTICES}
The renormalization group provides us with an additional point of view,
especially on hierarchical lattices.
Let us remember the following features of the renormalization group:
(i) The critical point is attracted toward the unstable fixed point.
(ii) The partition function does not change its functional form by the
renormalization on hierarchical lattices; only the values of arguments change.
Therefore the renormalized system also has a representative point
in the same space $(u_1(K),u_2(K),\cdots,u_n(K))$ as in Fig. \ref{Trajectory}.
The renormalization flow from the critical point $p_c$ reaches the fixed point C,
$(u^{(\infty)}_1,u^{(\infty)}_2,\cdots, u^{(\infty)}_n)$.
Here the superscript means the number of renormalization steps.
There is a point $d_c$ related to $p_c$ by the duality,
which is expect to also reach the same fixed point C since $p_c$ and $d_c$
represent the same critical point due to Eq. (\ref{PF2}).
Considering the above property of the renormalization flow as well as the duality, we find
that the duality relates two trajectories of the renormalization flow from $p_c$ and from $d_c$.
The same applies to the whole part of both curves, thin and dashed.
In other words, after a sufficient number of renormalization steps, the thin curve
representing the original system and the dashed curve for the dual system both
approach the common renormalized system depicted as the bold curve in Fig. \ref{Trajectory},
which goes through the fixed point C.

The partition function is then expected to become a single-variable function
along the bold curve.
This fact enables us to improve the method so that the exact location of the
multicritical point is obtained asymptotically, which can be given by
$x^{(s \to \infty)}_0(K) = {x_0^*}^{(s \to \infty)}(K)$. 
If we regard $x_0(K)={x_0^*}(K)$ as the zeroth approximation for the location of the
multicritical point, it is expected that $x^{(1)}_0(K)={x_0^*}^{(1)} (K)$ is the
first approximation and can lead to more precise results than $x_0(K)={x_0^*}(K)$ does.

Our method by the duality analysis in conjunction with the renormalization group indeed has
given the results in excellent agreement with the exact estimations within numerical
errors on several self-dual hierarchical lattices as summarized in Table \ref{Con}.
\begin{table}[htbp]
\begin{center}
\begin{tabular}{ccc}
\hline
$p_c$ (without RG) & $p_c$ (with RG) & $p_c$ (numerical) \\
\hline
$0.8900$ & $0.8920$ & $0.8915(6)$ \\
$0.8900$ & $0.8903$ & $0.8903(2)$ \\
$0.8900$ & $0.8892$ & $0.8892(6)$ \\
$0.8900$ & $0.8895$ & $0.8895(6)$ \\
$0.8900$ & $0.8891$ & $0.8890(6)$ \\
\hline
\end{tabular}
\end{center}
\caption{Comparison of the methods with and without RG and numerical estimations
for several self-dual hierarchical lattices \cite{ONB}.}
\label{Con}
\end{table}
\section{FURTHER DEVELOPMENTS }
\begin{figure}
\begin{center}
\includegraphics[width=90mm]{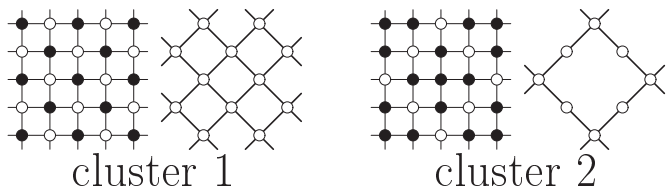}
\end{center}
\caption{The basic clusters used on the square lattice.
The spins marked black on the original lattice are traced out instead
of the iterative renormalization.}
\label{fig3}
\end{figure}

The above method has also been generalized to be applicable to Bravais lattices \cite{Ohzeki}.
Let us take an example of the square lattice.
Instead of the iterative renormalization, we consider to sum over a part of the spins,
to be called a cluster, on the square lattice as shown in Fig. \ref{fig3} to incorporate
many-body effects such as frustration inherent in spin glasses.
To this end, we define the principal Boltzmann factors $x_{0}^{(s)}$ and its dual
$x_{0}^{\ast (s)}$ as those with all spins surrounding the cluster in the up state.
We assume that a single equation gives the accurate location of the multicritical
point $x_{0}^{(s)}(K)=x_{0}^{\ast (s)}(K)$, where the superscript $s$ stands
for the type of the cluster.
Recent numerical investigations on the square lattice have given $p_c = 0.89081(7)$ \cite{Hasen}, $p_c = 0.89083(3)$ \cite{Toldin} and $p_c = 0.89061(6)$ \cite{Queiroz}, while the present method has estimated $p_c = 0.890725$
by cluster 1 of Fig. \ref{fig3}, and $p_c = 0.890822$ by cluster 2 \cite{Ohzeki}.
If we deal with clusters of larger sizes, the new method is expected to show systematic
improvements toward the exact answer from the point of view of renormalization.

The method of the renormalization group is applicable also away from the NL.
For example, the slope of the phase boundary at the pure ferromagnetic limit has
been estimated to be $1/T_c\times dT/dp \approx 3.2091\cdots$ on the square lattice
by perturbation \cite{Domany}.
This result is applicable also to any self-dual hierarchical lattices.
The present method with the renormalization group taken into account shows
that this is not the case. The result depends on the type of lattice,
e.g. $3.2786\cdots$ and $3.4390\cdots$ \cite{OH}.

\section{CONCLUSION}
The hierarchical lattices provide a very effective platform to test
new ideas as has been exemplified in the present study.
Investigations are notoriously hard for spin glasses on
finite-dimensional systems both analytically and numerically.
On hierarchical lattices, on the other hand, numerically exact
calculations can be carried out, and, in addition,
hierarchical lattices share many features with finite-dimensional
systems in contrast to mean-field systems.
Analytical methods can also be implemented with relative ease
on hierarchical lattices, which leads to the significant
improvements in the prediction of the location of the multicritical point.
Hierarchical lattices will continue to play key roles
in the studies of spin glass and other complex systems.

We thank financial supports by the CREST, JST.


\end{document}